\documentclass[aps,prl,twocolumn,groupedaddress]{revtex4}

\usepackage{graphicx}
\begin{document}


\title{Optimized Dynamical Decoupling in a Model Quantum Memory}



\author{Michael J. Biercuk}
\email[To whom correspondence should be addressed: ]{biercuk@boulder.nist.gov}
\altaffiliation{\emph{also} Georgia Inst. of Technology, Atlanta, Georgia}
\author{Hermann Uys}
\altaffiliation{\emph{also} Council for Scientific and Industrial Research, Pretoria, South Africa}
\author{Aaron P. VanDevender}
\author{Nobuyasu Shiga}
\altaffiliation{\emph{Present Address} NICT, Tokyo, Japan}
\author{Wayne M. Itano}
\author{John J. Bollinger}
\affiliation{NIST Time and Frequency Division, Boulder, CO}




\date{\today}

\begin{abstract}
\indent Any quantum system, such as those used in quantum information, magnetic resonance, or the like, is subject to random phase errors that can dramatically impact the fidelity of a desired quantum operation or measurement\cite{NC2000}.  In the context of quantum information, quantum error correction techniques have been developed to correct these errors, but resource requirements are extraordinary.  Realizing a physically tractable quantum information system will thus benefit significantly if qubit error rates are far below the so-called Fault-Tolerance error threshold\cite{NC2000}, predicted to be of order $10^{-3}-10^{-6}$ .  The need to realize such low error rates motivates a search for alternate strategies to suppress errors in quantum systems\cite{Zoller2005}.  We present experimental measurements on a model quantum system that demonstrate our ability to dramatically suppress qubit error rates by the application of optimized dynamical decoupling \cite{Haeberlen1976, Viola1998,Viola1999, Zanardi1999,Vitali1999, Byrd2003,Khodjasteh2005} pulse sequences in a variety of experimentally relevant noise environments.  We provide the first demonstration of an analytically derived pulse sequence developed by Uhrig \cite{Uhrig2007}, and find novel sequences through active, real-time experimental feedback.  These new sequences are specially tailored to maximize error suppression without the need for a priori knowledge of the ambient noise environment.  We compare these sequences against the Uhrig sequence, and the well established CPMG-style spin echo\cite{Vandersypen2004,Witzel2007}, demonstrating that our locally optimized pulse sequences outperform all others under test.  Numerical simulations show that our locally optimized pulse sequences are capable of suppressing errors by \emph{orders of magnitude} over other existing sequences.  Our work includes the extension of a treatment to predict qubit decoherence \cite{Uhrig2008,Cywinski2008} under realistic conditions, including the use of finite-duration, square $\pi$ pulses, yielding strong agreement between experimental data and theory for arbitrary pulse sequences.  These results demonstrate the robustness of qubit memory error suppression through dynamical decoupling techniques across a variety of qubit technologies\cite{Yao2007,Witzel2007, Lee2008,Zhang2008, Yang2008}.
\end{abstract}


\maketitle

\indent We consider classical phase randomization of our qubit due to the action of the environment as the dominant source of errors, a situation germane to many quantum systems.  Accordingly, we may write a Hamiltonian as $H=\frac{1}{2}[\Omega+\beta(t)]\sigma_{Z}$,
where $\Omega$ is the qubit splitting, $\beta$ is a classical random variable\cite{Kuopanportti2008,Cywinski2008}, and $\sigma_{Z}$ is a Pauli operator.  As in reference \cite{Cywinski2008}, we may write the time evolution of a superposition state initially oriented along $\hat{Y}$ under the influence of this Hamiltonian as $|\Psi(t)\rangle=\frac{1}{\sqrt{2}}(e^{-i\Omega t/2}e^{-\frac{i}{2}\int_{0}^{t}\beta(t')dt'}|\uparrow\rangle+e^{i\Omega t/2}e^{\frac{i}{2}\int_{0}^{t}\beta(t')dt'}i|\downarrow\rangle)$, with $|\uparrow\rangle$ and $|\downarrow\rangle$ the qubit states.  The term $\beta(t)$ adds a random phase to the time evolution of the state, corresponding to a random rotation around the equator of the Bloch sphere.  The buildup of such a random phase results in decoherence as an observer loses track of the Bloch vector in the equatorial plane.  However, the application of a $\pi$ pulse around $\hat{X}$ (henceforth denoted $\pi_{X}$), at time $t'=t/2$ will approximately time-reverse the state evolution, so long as fluctuations in $\beta$ are slow relative to the allowed free-precession time of the qubit.  This is the basis of the Hahn spin echo, a fundamental technique for preserving coherence in nuclear magnetic resonance and electron spin resonance systems\cite{Vandersypen2004}.
\\
\indent Given an arbitrary noise power spectrum $S_{\beta}(\omega)$, we would expect that the Hahn echo acts as a high-pass filter, mitigating phase errors associated with slowly varying Fourier components of $\beta$.  It was shown that this general interpretation can be extended to multipulse sequences.  Following references \cite{Uhrig2008} and \cite{Cywinski2008}, for any $n$ pulse sequence of total qubit evolution time $\tau$, the coherence of the state is given as $W(t)=|\overline{\langle\sigma_{Y}\rangle(t)}|=e^{-\chi(t)}$, where angled brackets indicate an expectation value, and the overline indicates an ensemble average in a frame rotating with frequency $\Omega$.  In this expression, \begin{equation}
\chi(t)=\frac{2}{\pi}\int\limits_{0}^{\infty}\frac{S_{\beta}(\omega)}{\omega^{2}}F(\omega t) d \omega,
\end{equation}
where the filter function $F(\omega\tau)$ contains all information about how the pulse sequence will preserve qubit coherence under the influence of $S_{\beta}(\omega)$.  $F(\omega\tau)$ is calculated from $F(\omega\tau)=|\tilde{y}_{n}(\omega\tau)|^{2}$, where $\tilde{y}_{n}(\omega\tau)$ is the Fourier transform of the time-domain filter function, $y_{n}(t)$ (Figs. 1b $\&$ 1c).  The time domain filter function alternates between +1 and -1 for successive free precession periods.
\\
\indent We begin by studying two distinct pulse sequences as will be described in order below: CPMG, and UDD (pulse spacings illustrated in Fig. 1a).  CPMG is an extension of the Hahn spin echo to a multipulse form, incorporating evenly spaced $\pi$ pulses about an axis rotated 90 degrees from the direction imparting the initial $(\frac{\pi}{2})_{X}$) (Fig 1d).  This sequence has been shown to be robust against a variety of phase and rotation errors, and does a particularly good job at refocussing the Bloch vector \cite{Haeberlen1976}.
\\
\begin{figure}
  \includegraphics[width=\columnwidth]{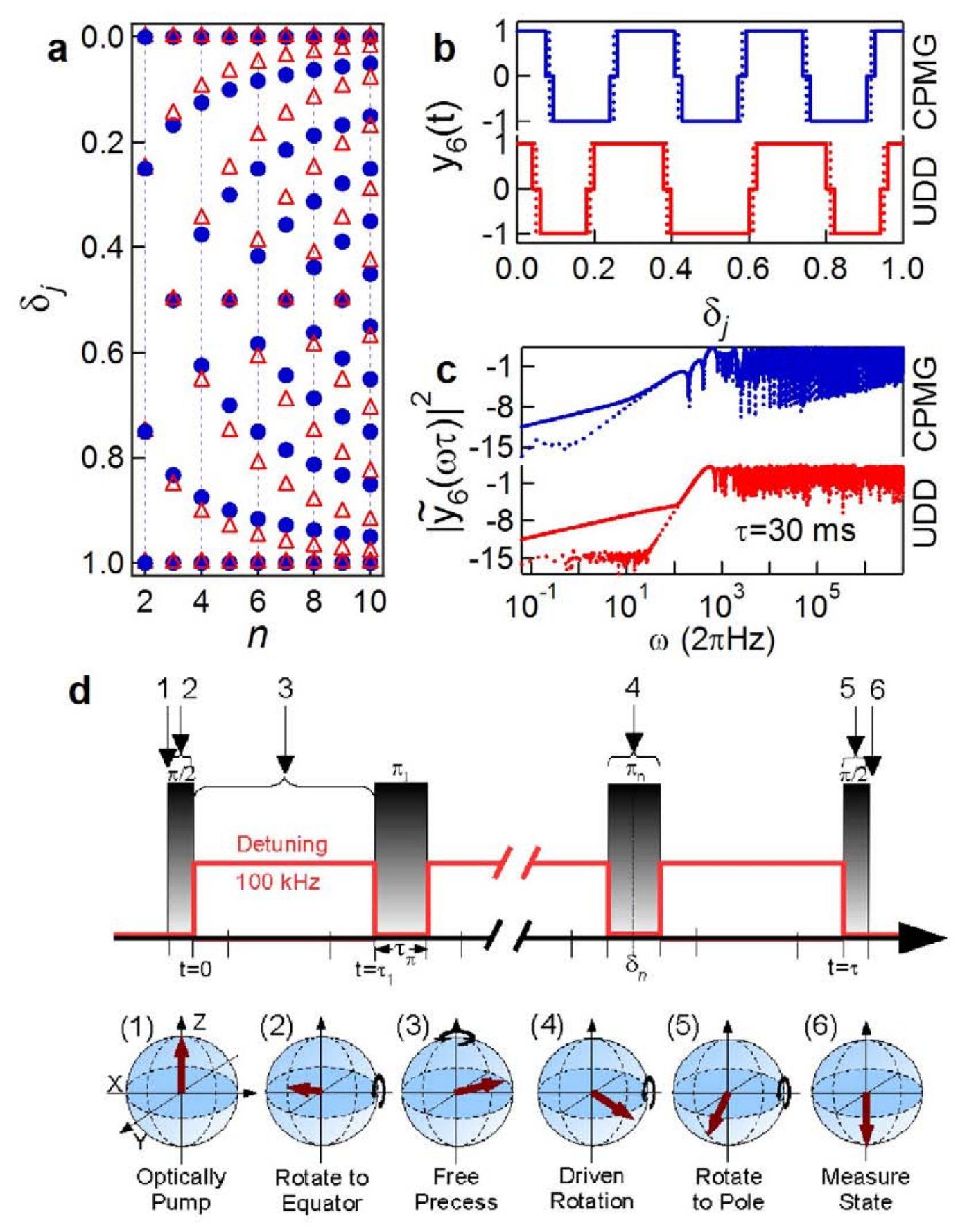}\\
    \caption{CPMG and UDD pulse sequence schematics. (a) Fractional pulse locations, $\delta_{j}$, of CPMG (solid markers) and UDD (open markers) sequences as a function of pulse number, $n$. (b) Examples of the time-domain filter function, $y_{6}(t)$, for the CPMG and UDD pulse sequences with 6 $\pi$-pulses.  The dotted line represents the time-domain filter function assuming delta-function $\pi$ pulses, while the solid line represents the time-domain filter function accounting for a finite $\tau_{\pi}$.  (c) Log-log plot of the log of the filter function, $F(\omega\tau)=|\tilde{y}_{6}(\omega\tau)|^{2}$, for sequence length $\tau=$30 ms, and $\tau_{\pi}=$185 $\mu$s.  Dotted lines indicate filter function with delta-function $\pi$ pulses, solid lines account for finite $\tau_{\pi}$.  (d) Generalized schematic of an experimental sequence showing key procedures and designations of key times.}
\end{figure}
\indent Uhrig discovered \cite{Uhrig2007} that for an $n$ pulse sequence, it is possible to modify the sharpness of the filter function, and hence the efficiency of noise suppression relative to CPMG, \emph{simply by changing the relative positions of the pulses within the sequence} (Figs. 1a $\&$ 1c).  He developed an optimized sequence for the spin boson model (appropriate for many solid-state systems), UDD, designed to increase the suppression of errors at short times --- the so-called ``high-fidelity'' regime (the UDD sequence was later shown to be general\cite{Lee2008,Cywinski2008,Yang2008}).  Pulses are no longer evenly spaced, and deviation from the CPMG sequence grows with $n$ (CPMG and UDD are equivalent for $n=2$).
\\
\indent The experimental noise spectrum can vary significantly between qubit implementations (e.g., semiconducting quantum dots vs. ultracold atoms), and as such, it is not sufficient to find and test a single pulse sequence that suppresses noise for only one characteristic spectrum.  For example, in superconducting qubit systems \cite{Nakamura2002,Faoro2006,Koch2007,Cywinski2008}, fluctuating electric charges and spin centers produce noise spectra varying as $1/\omega$.  By contrast, a spin-boson model for quantum noise in semiconducting quantum dots suggests the presence of noise with an ohmic spectrum, $S_{\beta}(\omega)\propto\omega$, and a sharp cutoff\cite{Leggett1987,Uhrig2008_2,Uhrig2008}. In order to test the efficacy of any pulse sequence, one must develop a method and testbed capable of exhibiting a variety of realistic noise environments.  In this manner we may think of the testbed as being a \emph{model} quantum memory capable of simulating nearly any other.
\\
\indent We realize such a model quantum memory in an array of $\sim$ 1000 $^{9}$Be$^{+}$ ions in a Penning ion trap\cite{Taylor2007}.  Previous experiments have demonstrated that under appropriate conditions these ions form 2D or 3D arrays with well defined crystal structure \cite{itaw98, mitchell98,huap98a} (Inset, Fig. 2a). This system mimics dense qubit arrays that might be constructed in a variety of physical incarnations including atomic systems, superconducting circuits, and semiconducting quantum dots.
\\
\indent  The qubit states are realized using a ground-state electron-spin-flip transition as shown in Fig. 2a (also see Methods).  Qubit operations are achieved by directly driving this $\sim$124 GHz transition via a quasi-optical microwave system (Fig. 2b).  We drive $\sigma_{X}$ rotations by the application of square microwave pulses to the qubit (Figs. 1d, 2c, $\&$ 2d), while controlled $\sigma_{Z}$ rotations are achieved by detuning the microwaves from resonance and allowing the qubit state to freely precess for a precise time period (Fig. 2e).  These methods are particularly well-suited to dynamical decoupling studies compared to laser-mediated qubit rotations due to the absence of spontaneous emission in the microwave/millimeter-wave regime.  Doppler cooling of ion motion along the axis of the Penning trap\cite{jenm05} using UV laser light red-detuned from an atomic transition yields ion temperatures of order 1 mK.  State initialization occurs via optical pumping (Fig. 2a), and state readout is achieved by fluorescence detection on the same cycling transition used for cooling \cite{brel88}.  We are able to initialize the system in a pure state with high fidelity, and perform a strong projective measurement, despite the use of ensemble techniques\cite{itaw93}.
\\
\begin{figure}
\includegraphics[width=\columnwidth]{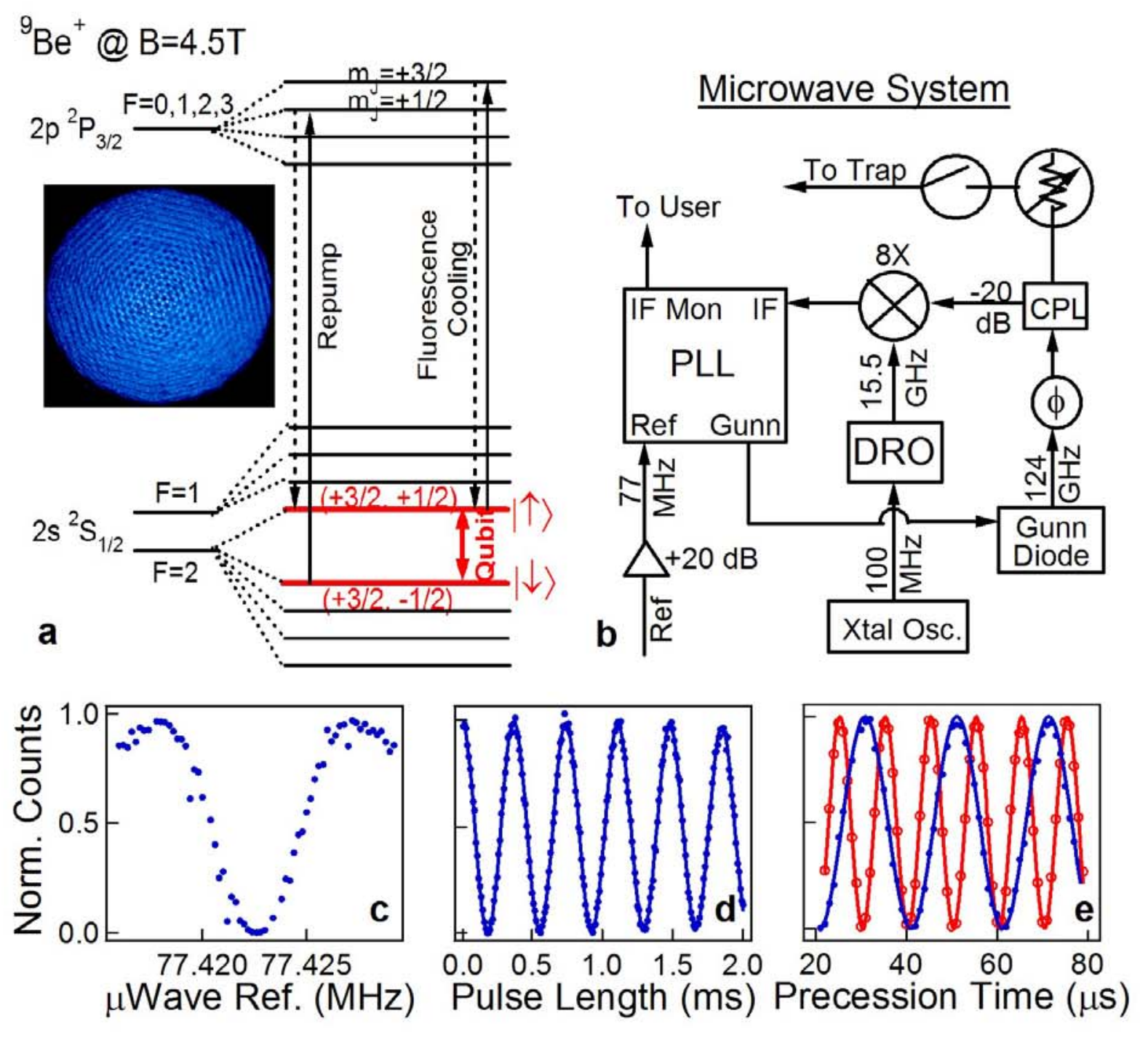}\\
\caption{ $^{9}$Be$^{+}$ qubit structure and coherent control.  (a) Relevant atomic structure of $^{9}$Be$^{+}$ at 4.5 T.  Qubit states labeled with (m$_{I}$,m$_{J}$), the nuclear and total spin projections of the atom along the quantization axis.  (Inset) Strobed optical image of ion fluorescence showing hexagonal-close-packed order with spacing $\sim$10 $\mu$m.  (b) Schematic block diagram of the microwave system used to drive qubit rotations.  PLL = Phase Lock Loop, DRO = Dielectric Resonator Oscillator, CPL = Coupler.  77 MHz PLL reference provided by computer-controlled DDS or the beat note of DDS and a frequency modulated synthesizer during noise injection (see Methods).  (c) Qubit transition driven via a square microwave $\pi$ pulse of $\tau_{\pi}\approx185$ $\mu$s.  Readout consists of fluorescence detection immediately before and after application of microwave pulses followed by normalization to account for slow fluctuations in laser intensity (Normalized Counts).  Each data point consists of the average of 20 experiments.  (d) Rabi oscillations driven on-resonance give a decay time of 30-40 ms.  $\tau_{\pi}$ is tunable via an input attenuator. (e) Ramsey fringes measured for two different microwave detunings, 50 kHz and 100 kHz ($\pi/2$ pulses were driven on-resonance and the microwaves detuned during the free-precession period).  Detuning was controlled by switching between pre-programmed frequency output profiles via a TTL input on the DDS.}
\end{figure}
\indent We apply pulse sequences consisting of a few to more than 1000 $\pi_{X}$ pulses.  We have successfully extended our qubit coherence time (i.e. $1/e$ decay time) from approximately 1 ms as measured via Ramsey free-induction decay, to over 200 ms using 500 $\pi_{X}$ pulses in a CPMG sequence.  We find that for large $n$, our minimum measured deviation from full projection to $|\downarrow\rangle$ is of order 10 $\%$ (likely due to long-time amplitude instability of our microwave system).  Accordingly, we focus primarily on sequences with $n\leq10$, allowing us to compare pulse sequences in a regime where the minimum error rate is $\leq$1 $\%$.
\\
\indent We apply the CPMG and UDD sequences for various pulse numbers (Figs. 1a $\&$ 1d), and measure ensemble-averaged state decoherence due to ambient magnetic field fluctuations as a function of total free-precession time, as shown in Fig. 3a (see Methods).  The ambient magnetic field noise in our high-field superconducting magnet is measured directly, varying approximately as $1/\omega^{2}$ ($S_{\beta}(\omega)\propto1/\omega^{4}$) with additional sharp spurs of undetermined origin, including a prominent feature at $\sim$153 Hz.  CPMG appears to outperform the analytically optimized UDD sequence for all $n\leq10$ in this noise environment, which has a soft high-frequency cutoff \cite{Uhrig2008}.  Further, the data in Fig. 3a demonstrate that it is possible to extend the qubit coherence time by adding $\pi$ pulses, as expected.
\\
\indent Fitting our experimental data requires that we account for finite $\pi$-pulse durations in expressions for the pulse sequence filter function, diverging from the zero-pulse-length assumptions made in most literature on dynamical decoupling\cite{Lee2008,Cywinski2008, Uhrig2008_3}.  We assume that dephasing is negligible during the application of a $\pi_{X}$ pulse, based on the observation of Rabi-flopping decay times more than an order of magnitude longer than Ramsey free-induction decay times, and build on the theoretical descriptions of Uhrig\cite{Uhrig2008} and Cywinski\cite{Cywinski2008}.  The above assumption leads to the insertion of a delay, $\tau_{\pi}$, between each free-precession time, during which the filter function in the time domain has value zero (rather than $\pm1$ as in \cite{Cywinski2008, Uhrig2007, Uhrig2008_3}).  Moving to the frequency domain, we may write the filter function of an arbitrary $n$-pulse sequence as
\begin{eqnarray}
&F(\omega\tau)=|\tilde{y}_n(\omega\tau)|^{2}\nonumber\\
&=|1+(-1)^{n+1}e^{i\omega\tau}\\
&\mbox{}+2\sum\limits_{j=1}^n(-1)^je^{i\delta_j\omega\tau}\cos{\left(\phi_\pi \omega\tau/2\right)}|^{2} \nonumber,
\end{eqnarray}
where $\delta_{j}\tau$ is the time of the center of the $j^{\rm th}$ $\pi_{X}$ pulse, $\tau$ is the sum of the total free-precession time and $\pi$-pulse times, and $\phi_\pi$ is the ratio of $\tau_{\pi}$ compared to the total sequence length $\tau$ (Figs. 1b, 1c $\&$ 1d).  To this order of approximation, all information pertaining to finite pulse lengths is accounted for by the simple addition of a cosine term in the equation.
\\
\indent Fits to experimental data show good agreement with theory.  In Fig. 3a, the free fit parameters are the overall noise strength and the relative strength of the 153 Hz spur in our noise spectrum.  This spur can be observed to slowly change amplitude in real-time, and is entirely responsible for the plateau-like feature we see in our decoherence curves.  Increasing the strength of this spectral feature changes the plateau-like feature to a rounded hill of increasing height.  We believe that deviations between our experimental data and fitting functions are dominated by slow as well as discontinuous changes in the ambient noise environment.
\\
\begin{figure}
\includegraphics[width=200pt]{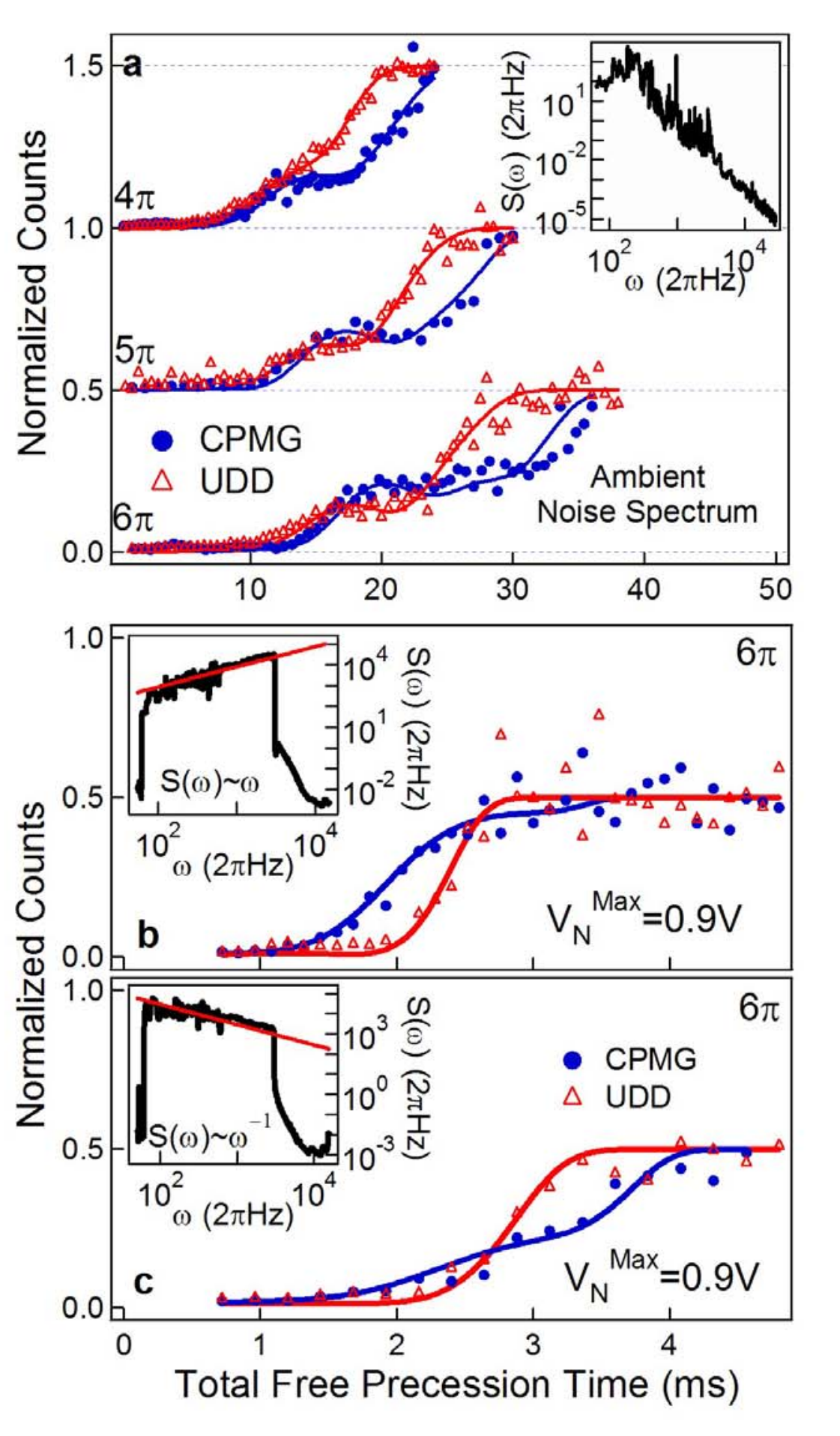}\\
\caption{Pulse sequence performance in the presence of various noise spectra.  (a) Decoherence traces as a function of free-precession time for CPMG and UDD and ambient magnetic field noise.  Phase errors are manifested as nonzero fluorescence detection.  Traces saturate at 0.5 for total phase randomization.  Traces for $n$=4,5 offset by 0.5 units for clarity.  Each data point corresponds to 50 averages.  Fits utilize ambient noise spectrum (see inset) and two free parameters - overall noise scaling ($\alpha$) and scaling factor for the noise spur at $\sim$153 Hz ($\gamma$).  Extracted fit parameters yield $\alpha=1.46\pm0.31$ and $\gamma=0.23\pm0.04$.  Individual fit uncertainty is $\sim$0.05-0.07 for each trace.  Variations in extracted noise intensities are consistent with direct observation of slow fluctuations in the noise spectrum, including significant changes in the amplitude of the $\sim$153 Hz spur.  (Inset)  Ambient noise spectrum measured via a solenoid embedded in our NMR magnet.  (b) Performance of UDD and CPMG for Ohmic spectrum with 500 Hz cutoff (see inset).  UDD outperforms CPMG by increasing factors as the noise intensity is increased.  Single fit parameter: noise scaling, $\alpha$.  Extracted noise scaling: $\alpha_{\rm Ohmic}=1.49\pm0.19$, expected $\alpha=$1.65 for this noise intensity.  (c) Sequence performance in a $1/\omega$ noise spectrum with a sharp cutoff.  Extracted noise scaling: $\alpha_{1/\omega}=3.32\pm0.15$.  Insets to (b) and (c)) Smoothed, measured noise on a log-log plot, measured using $V_{N}=0.7$ V.  Red line represents intended noise envelope up to high-frequency cutoff.  Measured noise spectra extending from .01 Hz to 1 MHz used for fitting, but truncated for visibility.}
\end{figure}
\indent Numerical simulations suggest that in the ``high-fidelity regime,'' UDD is capable of significantly outperforming CPMG even in this noise environment, suppressing errors by several orders of magnitude to yield ultimate fidelities in excess of 99.99 $\%$ \cite{Uhrig2007, Cywinski2008}.  In order to emphasize the differences between these sequences, and overcome limitations in measurement fidelity, we artificially inject noise to simulate systems where UDD outperforms CPMG in the lower-fidelity regime (see Methods).
\\
\indent We inject noise with ohmic and $1/\omega$ power spectra and test the relative performance of UDD and CPMG (Figs. 3b $\&$ 3c).  Our data indicate that the UDD sequence dramatically outperforms CPMG in the presence of noise with an ohmic spectrum and a sharp cutoff\cite{Uhrig2007} --- a significant departure from results under ambient noise.  By contrast, using a $1/\omega$ power spectrum with a sharp cutoff, we find that over the entire range of accessible noise intensities, CPMG performs similarly to UDD.  As expected, peak noise power at low-frequencies is well-filtered by both sequences, yielding longer overall coherence times for the $1/\omega$ spectrum relative to the application of an Ohmic spectrum, and consistent with theoretical work \cite{Uhrig2008}.  Fits to theoretical expressions for the UDD and CPMG filter functions incorporating the effects of finite $\tau_{\pi}$ show strong agreement using a single fitting parameter: $\alpha$, a scaling factor for the overall noise level.
\\
\indent These experimental measurements are reproducible between different ion clouds, and do not appear to vary significantly with changes in ion number.  Additionally, the UDD sequence appears robust to rounding errors associated with our pulse timing accuracy of 50 ns, and performs similarly for all input states on the equator of the Bloch sphere (not shown).  We often used evenly spaced pulses (Periodic Dynamical Decoupling\cite{Viola1998}) as a benchmark; in all cases it performs worst of the sequences under consideration.  Finally, in all noise environments and for all pulse sequences, neglecting the effects of finite $\tau_{\pi}$ dramatically diminishes our fit quality and leads to wide variations in the extracted noise strength, especially in regimes where the free-precession period is comparable to or less than $\tau_{\pi}$.
\\
\indent We extend the ideas of Uhrig by developing novel Locally Optimized Dynamical Decoupling (LODD) pulse sequences that are tailored to a given experimental noise environment and a fixed free precession time.  We employ the Nelder-Mead simplex method for optimization in an $n$-dimensional space, manipulating the relative pulse positions in the sequence.  In our numerical simulations the Nelder-Mead method converges to an optimized pulse sequence for given initial conditions in less than 100 iterations, providing a suite of locally optimized pulse sequences (Inset, Fig. 4a).
\\
\begin{figure}
\includegraphics[width=200pt]{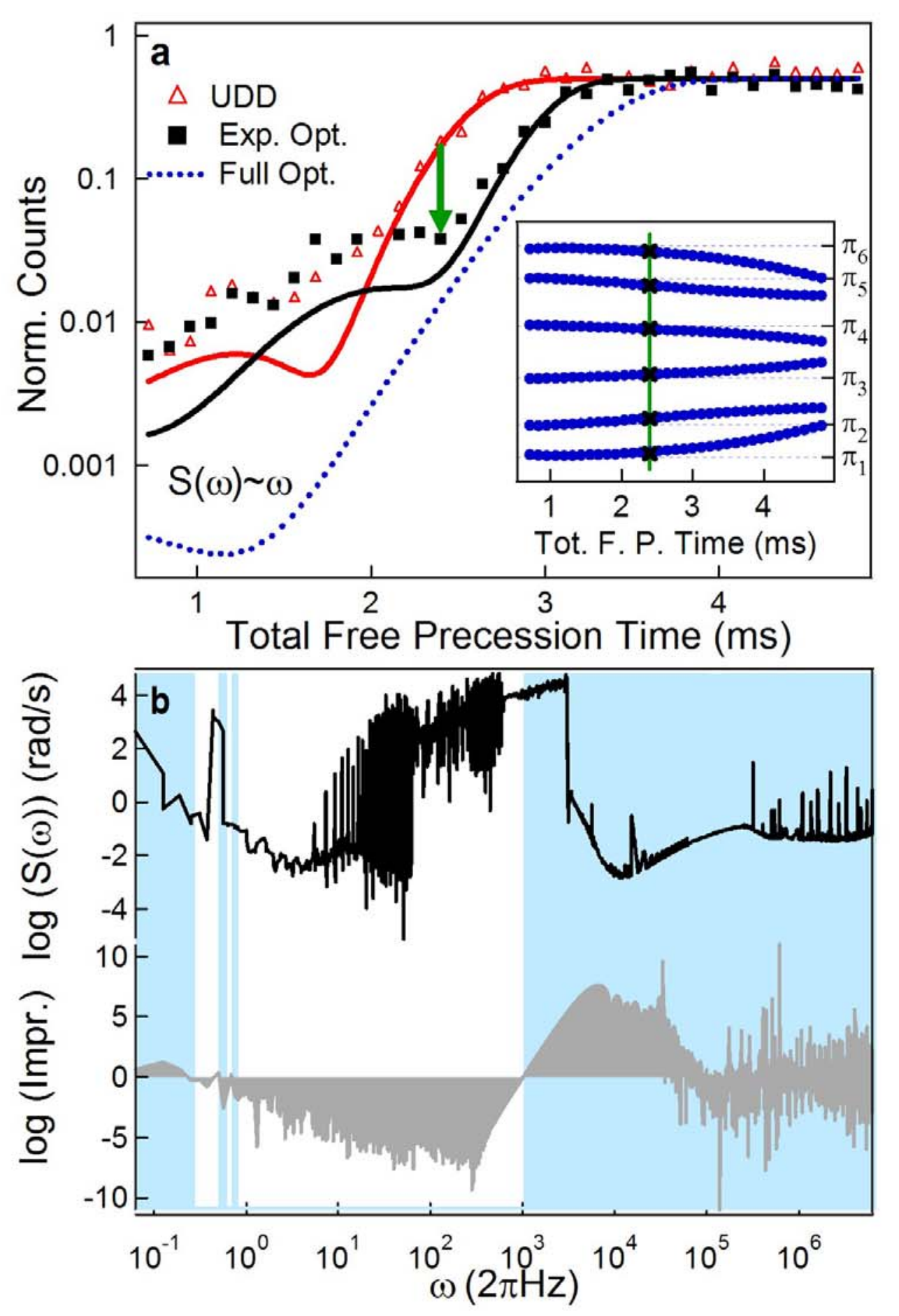}\\
\caption{Nelder Mead pulse-sequence optimization using the ohmic spectrum in Fig. 3b.  (a) Experimental measurements of 6$\pi$-pulse UDD and experimentally optimized pulse sequence derived from UDD on a logarithmic scale.  Fit parameters:  $\alpha_{UDD}=0.69$, $\alpha_{Exp. Opt.}=0.77$.  Green arrow indicates the total sequence length for which experimental optimization was performed.  The measured error rate using the LODD sequence is $\sim7\times$ lower than that obtained using CPMG.  Blue dotted line indicates numerical simulation of decoherence with the same measured noise spectrum, $\alpha=0.77$, and with the pulse sequence optimized for each total sequence length, $\tau$.  (Inset) Optimized sequence as a function of total free-precession time (solid markers).  Black X's indicate the $\pi$ pulse positions of the single experimentally optimized sequence.  Ticks and horizontal grids correspond to original UDD pulse positions.  (b) Filtration benefits of experimentally optimized pulse sequence tailored to injected noise environment.  Noise power spectrum (upper trace) and logarithmic improvement of optimized filter function relative to UDD filter function (lower trace).  Positive numbers (also background shading) indicate regions of the noise spectrum where the optimized sequence provides noise suppression superior to UDD.}
\end{figure}
\indent The efficacy of any optimization method in determining LODD pulse sequences, however, is limited by the degree to which $S_{\beta}(\omega)$ is known.  In addition to general noise models, the specific measurement hardware and laboratory environment will add frequency components to $S_{\beta}(\omega)$ that are not easily measured or predicted by theory.  We therefore modify the Nelder-Mead search algorithm by replacing numerical calculation inputs with the results of experimental measurements, resulting in a form of real-time, active, experimental feedback for quantum control \cite{Weinacht99}.  This extension allows us to find optimized pulse sequences \emph{without definite knowledge of the experimental noise environment}.
\\
\indent Experimental optimization results for an Ohmic spectrum are displayed in Fig. 4a, along with a schematic depiction of the resultant optimized pulse sequence compared with UDD (inset).  We find that small modifications in the pulse positions (for both ohmic and $1/\omega$ spectra) produce significant improvements for the selected total free-precession time, and fits to theoretical expressions for qubit decoherence yield strong agreement.  In the data presented we are able to suppress the qubit error rate by nearly an order of magnitude over CPMG by the use of a LODD pulse sequence (a factor of five relative to the starting sequence, UDD).  We also show via numerical simulation that it is possible to continuously vary the pulse sequences such that the filter function is optimized for all total free precession times (blue dashed line, Fig. 4a), yielding sequences which outperform CPMG \emph{and} UDD in the high-fidelity regime by an order of magnitude.
\\
\indent In summary, we have demonstrated the efficacy of phase error suppression via optimized dynamical decoupling pulse sequences applied to a model quantum memory.  This experimental system has been employed to test various pulse sequences under experimentally realistic noise environments, yielding good agreement with theoretical predictions for qubit decoherence under sequence application.  Finally, we have developed a real-time active feedback technique to experimentally produce locally optimized pulse sequences without any knowledge of the experimental noise environment, making these methods generally useful to any experimental setting.  Future studies will incorporate the concatenation of optimized pulse sequences\cite{Khodjasteh2005, Uhrig2008_2}, and will examine the use of pulse-shaping for further error reduction.  Our results provide key ingredients of a quantum toolkit which will make the production of a functional and useful quantum computer more realistic.

\section{methods}
\subsection{Experimental control}
\indent In our experiments we control several parameters using a centralized programmable-logic-device (PLD)-based pulse generator.  We have pulsed control over detection/cooling and repump lasers,  pulse counters, microwave reference frequency, and a microwave shutter which provides $\sim$25 dB of isolation in the off state.  All relevant pulse sequences are generated in software on a controlling PC and sent to the PLD.  Similarly, various settings are sent directly to the DDS which serves as a reference for our microwave system, and to the data acquisition system which records fluorescence detection events from a phototube.   An operator command launches the experiment which is synchronized to main line frequency and triggers the PLD to begin its pulse sequence.  A complete experiment including state initialization, measurement, and the application of the pulse sequence itself takes approximately 200ms.  The experiment is conducted a fixed number of times and the results recorded and averaged before new experimental conditions are uploaded.
\subsection{Pulse sequence application}
\indent An experiment begins with initialization of the qubit in  $|\uparrow\rangle$ (Fig. 1d).  The qubits are then rotated to the equator using a $\frac{\pi}{2}$ pulse around the $\hat{X}$ axis, $(\frac{\pi}{2})_{X}$, forming a superposition state along $\hat{Y}$.  The microwaves are detuned during the interpulse period, and the qubits are allowed to precess about $\hat{Z}$ at a rate controlled by the detuning from resonance (see Figs. 1d $\&$ 2e).  Detuning the microwaves during the interpulse period also mitigates the effect of microwave leakage due to the $\sim$25 dB isolation of our microwave switch.  Free precession is punctuated by the application of on-resonance $\pi_{X}$ pulses comprising the sequence under study.  At the end of an experimental cycle the qubit state is rotated such that in the absence of dephasing, the state evolves to $|\downarrow\rangle$ and is subsequently read-out.  By precisely controlling the length of time during which the microwaves are detuned during the first free-precession period (i.e., between the $(\frac{\pi}{2})_{X}$ and the first $\pi_{X}$), we initialize the qubit in any state on the equatorial plane of the Bloch sphere.  Control over the final precession period is set to project the qubits appropriately to the dark state (in the absence of dephasing).
\subsection{Measurement and operation fidelity}
\indent Using a low-dark-count phototube we achieve a state measurement fidelity of $\sim$99.5 $\%$, limited predominantly by stray-light scatter from trap electrodes (Fig. 2c).  We have microwave amplitude stability $>$99.9 $\%$, and our pulse timing resolution of 50 ns corresponds to a systematic pulse infidelity of $<$4$\times10^{-4}$.  A $\pi$ rotation is sensitive to these inaccuracies only to second order.  Careful monitoring of the resonance frequency allows us to maintain long-term microwave frequency stability relative to the qubit transition of $\sim1\times10^{-9}$ over any measurement run, despite slow drifts and occasional sudden shifts in the magnetic field (hence $\Omega$).  From the decay of Ramsey fringes we find that the integrated ambient noise spectrum yields a shot-to-shot magnetic field variation $\delta B/B\sim 7\times10^{-10}$ for 2 ms Ramsey experiments separated by $\sim$200 ms. Magnetic field gradients across the ion array produced comparable shifts in $\delta B$, yielding a $\pi$-pulse infidelity of $<$1 $\%$.
\subsection{Noise injection}
\indent We simulate the noise environment of other quantum systems by directly injecting frequency noise in our microwave phase-lock loop.  Noisy modulation of the microwave frequency is equivalent to external field fluctuations that modulate the qubit splitting, $\Omega$, relative to a fixed microwave drive.  By applying the noise only during the interpulse precession period we maintain high-fidelity $\pi_{X}$-rotations that otherwise would require shorter $\pi_{X}$-pulses.  This technique simulates systems that apply very strong and narrow $\pi_{X}$-pulses, similar to the original prescriptions for dynamical decoupling.
\\
\indent This process begins with the numerical generation of a desired noise spectrum;  all frequency components up to a given cutoff are provided a uniformly distributed random phase (between 0 and $2\pi$) and weighted by an envelope function dictated by the desired noise spectrum (e.g., $1/\omega$). These components are then Fourier transformed to produce a time trace whose ensemble-averaged two-time correlation function reproduces the spectrum of interest.  An arbitrary waveform generator outputs this time trace to the external frequency modulation port of a synthesizer whose beat note with a computer-controlled DDS produces the reference frequency for our Gunn Diode phase-lock loop. A peak-peak modulation $V_{N}^{max}=$ 1 V corresponds to a maximum relative frequency error of $\sim$9$\times10^{-8}$ in a single time trace --- about twice the linewidth of our qubit transition --- and a time-averaged deviation of $\sim$300 Hz at the spectral peak.  We directly measure the noise power spectrum of our frequency-modulated carrier from 0.01-10$^{6}$ Hz using a phase-noise detection system, producing the spectra displayed in the insets of Fig. 3.
\subsection{Nelder Mead optimization}
\indent Experimental Nelder-Mead optimization produces LODD pulse sequences tailored to the user-input requirements of pulse number and total free-precession period.  Optimization begins either with PDD or the sequence performing best in a given noise environment.  The algorithm proceeds until changes in pulse spacing converges to values less than the timing resolution of our pulse generator, generally in about 50 iterations. Optimization for $n=4$, beginning from evenly spaced pulses in the ambient noise environment, yields a sequence similar to CPMG.  We are unable to improve significantly beyond CPMG, consistent with numerical simulations using noise with a soft high-$\omega$ cutoff. Optimization for a $1/\omega$ noise spectrum yielded a roughly 4$\times$ improvement in sequence performance over CPMG at the optimization point.

\bibliography{scibib}

\begin{thebibliography}{33}
\expandafter\ifx\csname natexlab\endcsname\relax\def\natexlab#1{#1}\fi
\expandafter\ifx\csname bibnamefont\endcsname\relax
  \def\bibnamefont#1{#1}\fi
\expandafter\ifx\csname bibfnamefont\endcsname\relax
  \def\bibfnamefont#1{#1}\fi
\expandafter\ifx\csname citenamefont\endcsname\relax
  \def\citenamefont#1{#1}\fi
\expandafter\ifx\csname url\endcsname\relax
  \def\url#1{\texttt{#1}}\fi
\expandafter\ifx\csname urlprefix\endcsname\relax\def\urlprefix{URL }\fi
\providecommand{\bibinfo}[2]{#2}
\providecommand{\eprint}[2][]{\url{#2}}

\bibitem[{\citenamefont{Nielsen and Chuang}(2000)}]{NC2000}
\bibinfo{author}{\bibfnamefont{M.}~\bibnamefont{Nielsen}} \bibnamefont{and}
  \bibinfo{author}{\bibfnamefont{I.}~\bibnamefont{Chuang}},
  \emph{\bibinfo{title}{Quantum Computation and Quantum Information}}
  (\bibinfo{publisher}{Cambridge University Press},
  \bibinfo{address}{Cambridge, England}, \bibinfo{year}{2000}).

\bibitem[{\citenamefont{Zoller et~al.}(2005)\citenamefont{Zoller, Beth, Binosi,
  Blatt, Briegel, Bruss, Calarco, Cirac, Deutsch, Eisert et~al.}}]{Zoller2005}
\bibinfo{author}{\bibfnamefont{P.}~\bibnamefont{Zoller}},
  \bibinfo{author}{\bibfnamefont{T.}~\bibnamefont{Beth}},
  \bibinfo{author}{\bibfnamefont{D.}~\bibnamefont{Binosi}},
  \bibinfo{author}{\bibfnamefont{R.}~\bibnamefont{Blatt}},
  \bibinfo{author}{\bibfnamefont{H.}~\bibnamefont{Briegel}},
  \bibinfo{author}{\bibfnamefont{D.}~\bibnamefont{Bruss}},
  \bibinfo{author}{\bibfnamefont{T.}~\bibnamefont{Calarco}},
  \bibinfo{author}{\bibfnamefont{J.}~\bibnamefont{Cirac}},
  \bibinfo{author}{\bibfnamefont{D.}~\bibnamefont{Deutsch}},
  \bibinfo{author}{\bibfnamefont{J.}~\bibnamefont{Eisert}},
  \bibnamefont{et~al.}, \bibinfo{journal}{European Physical Journal D}
  \textbf{\bibinfo{volume}{36}}, \bibinfo{pages}{203} (\bibinfo{year}{2005}).

\bibitem[{\citenamefont{Haeberlen}(1976)}]{Haeberlen1976}
\bibinfo{author}{\bibfnamefont{U.}~\bibnamefont{Haeberlen}},
  \emph{\bibinfo{title}{High Resolution NMR in Solids, Advances in Magnetic
  Resonance Series}} (\bibinfo{publisher}{Academic Press},
  \bibinfo{address}{New York}, \bibinfo{year}{1976}).

\bibitem[{\citenamefont{Viola and Lloyd}(1998)}]{Viola1998}
\bibinfo{author}{\bibfnamefont{L.}~\bibnamefont{Viola}} \bibnamefont{and}
  \bibinfo{author}{\bibfnamefont{S.}~\bibnamefont{Lloyd}},
  \bibinfo{journal}{Phys. Rev. A} \textbf{\bibinfo{volume}{58}},
  \bibinfo{pages}{2733} (\bibinfo{year}{1998}).

\bibitem[{\citenamefont{Viola et~al.}(1999)\citenamefont{Viola, Knill, and
  Lloyd}}]{Viola1999}
\bibinfo{author}{\bibfnamefont{L.}~\bibnamefont{Viola}},
  \bibinfo{author}{\bibfnamefont{E.}~\bibnamefont{Knill}}, \bibnamefont{and}
  \bibinfo{author}{\bibfnamefont{S.}~\bibnamefont{Lloyd}},
  \bibinfo{journal}{Phys. Rev. Lett.} \textbf{\bibinfo{volume}{82}},
  \bibinfo{pages}{2417} (\bibinfo{year}{1999}).

\bibitem[{\citenamefont{Zanardi}(1999)}]{Zanardi1999}
\bibinfo{author}{\bibfnamefont{P.}~\bibnamefont{Zanardi}},
  \bibinfo{journal}{Physics Letters} \textbf{\bibinfo{volume}{258}},
  \bibinfo{pages}{77} (\bibinfo{year}{1999}).

\bibitem[{\citenamefont{Vitali and Tombesi}(1999)}]{Vitali1999}
\bibinfo{author}{\bibfnamefont{D.}~\bibnamefont{Vitali}} \bibnamefont{and}
  \bibinfo{author}{\bibfnamefont{P.}~\bibnamefont{Tombesi}},
  \bibinfo{journal}{Phys. Rev. A} \textbf{\bibinfo{volume}{59}},
  \bibinfo{pages}{4178} (\bibinfo{year}{1999}).

\bibitem[{\citenamefont{Byrd and Lidar}(2003)}]{Byrd2003}
\bibinfo{author}{\bibfnamefont{M.}~\bibnamefont{Byrd}} \bibnamefont{and}
  \bibinfo{author}{\bibfnamefont{D.}~\bibnamefont{Lidar}},
  \bibinfo{journal}{Phys. Rev. A} \textbf{\bibinfo{volume}{67}},
  \bibinfo{pages}{012324} (\bibinfo{year}{2003}).

\bibitem[{\citenamefont{Khodjasteh and Lidar}(2005)}]{Khodjasteh2005}
\bibinfo{author}{\bibfnamefont{K.}~\bibnamefont{Khodjasteh}} \bibnamefont{and}
  \bibinfo{author}{\bibfnamefont{D.}~\bibnamefont{Lidar}},
  \bibinfo{journal}{Phys. Rev. Lett.} \textbf{\bibinfo{volume}{95}},
  \bibinfo{pages}{180501} (\bibinfo{year}{2005}).

\bibitem[{\citenamefont{Uhrig}(2007)}]{Uhrig2007}
\bibinfo{author}{\bibfnamefont{G.}~\bibnamefont{Uhrig}},
  \bibinfo{journal}{Phys. Rev. Lett.} \textbf{\bibinfo{volume}{98}},
  \bibinfo{pages}{100504} (\bibinfo{year}{2007}).

\bibitem[{\citenamefont{Vandersypen and Chuang}(2004)}]{Vandersypen2004}
\bibinfo{author}{\bibfnamefont{L.}~\bibnamefont{Vandersypen}} \bibnamefont{and}
  \bibinfo{author}{\bibfnamefont{I.}~\bibnamefont{Chuang}},
  \bibinfo{journal}{Rev. Mod. Phys.} \textbf{\bibinfo{volume}{76}},
  \bibinfo{pages}{1037} (\bibinfo{year}{2004}).

\bibitem[{\citenamefont{Witzel and Sarma}(2007)}]{Witzel2007}
\bibinfo{author}{\bibfnamefont{W.}~\bibnamefont{Witzel}} \bibnamefont{and}
  \bibinfo{author}{\bibfnamefont{S.~D.} \bibnamefont{Sarma}},
  \bibinfo{journal}{Phys. Rev. Lett.} \textbf{\bibinfo{volume}{98}},
  \bibinfo{pages}{077601} (\bibinfo{year}{2007}).

\bibitem[{\citenamefont{Uhrig}(2008{\natexlab{a}})}]{Uhrig2008}
\bibinfo{author}{\bibfnamefont{G.}~\bibnamefont{Uhrig}}, \bibinfo{journal}{New
  Journal of Physics} \textbf{\bibinfo{volume}{10}}, \bibinfo{pages}{083024}
  (\bibinfo{year}{2008}{\natexlab{a}}).

\bibitem[{\citenamefont{Cywinski et~al.}(2008)\citenamefont{Cywinski, Lutchyn,
  Nave, and Sarma}}]{Cywinski2008}
\bibinfo{author}{\bibfnamefont{L.}~\bibnamefont{Cywinski}},
  \bibinfo{author}{\bibfnamefont{R.}~\bibnamefont{Lutchyn}},
  \bibinfo{author}{\bibfnamefont{C.}~\bibnamefont{Nave}}, \bibnamefont{and}
  \bibinfo{author}{\bibfnamefont{S.~D.} \bibnamefont{Sarma}},
  \bibinfo{journal}{Phys. Rev. B} \textbf{\bibinfo{volume}{77}},
  \bibinfo{pages}{174509} (\bibinfo{year}{2008}).

\bibitem[{\citenamefont{Yao et~al.}(2007)\citenamefont{Yao, Liu, and
  Sham}}]{Yao2007}
\bibinfo{author}{\bibfnamefont{W.}~\bibnamefont{Yao}},
  \bibinfo{author}{\bibfnamefont{R.-B.} \bibnamefont{Liu}}, \bibnamefont{and}
  \bibinfo{author}{\bibfnamefont{L.}~\bibnamefont{Sham}},
  \bibinfo{journal}{Phys. Rev. Lett.} \textbf{\bibinfo{volume}{98}},
  \bibinfo{pages}{077602} (\bibinfo{year}{2007}).

\bibitem[{\citenamefont{W.Zhang et~al.}(2008)\citenamefont{W.Zhang,
  Kostantinidis, V.V, Dobrovitski, Harmon, Santos, and Viola}}]{Zhang2008}
\bibinfo{author}{\bibnamefont{W.Zhang}},
  \bibinfo{author}{\bibfnamefont{N.}~\bibnamefont{Kostantinidis}},
  \bibinfo{author}{\bibnamefont{V.V}},
  \bibinfo{author}{\bibnamefont{Dobrovitski}},
  \bibinfo{author}{\bibfnamefont{B.}~\bibnamefont{Harmon}},
  \bibinfo{author}{\bibfnamefont{L.}~\bibnamefont{Santos}}, \bibnamefont{and}
  \bibinfo{author}{\bibfnamefont{L.}~\bibnamefont{Viola}},
  \bibinfo{journal}{Phys. Rev. B} \textbf{\bibinfo{volume}{77}},
  \bibinfo{pages}{125336} (\bibinfo{year}{2008}).

\bibitem[{\citenamefont{Yang and Liu}(2008)}]{Yang2008}
\bibinfo{author}{\bibfnamefont{W.}~\bibnamefont{Yang}} \bibnamefont{and}
  \bibinfo{author}{\bibfnamefont{R.}~\bibnamefont{Liu}},
  \bibinfo{journal}{Phys. Rev. Lett.} \textbf{\bibinfo{volume}{101}},
  \bibinfo{pages}{180403} (\bibinfo{year}{2008}).

\bibitem[{\citenamefont{Kuopanportti et~al.}(2008)\citenamefont{Kuopanportti,
  Mottonen, Bergholm, Saira, Zhang, and Whaley}}]{Kuopanportti2008}
\bibinfo{author}{\bibfnamefont{P.}~\bibnamefont{Kuopanportti}},
  \bibinfo{author}{\bibfnamefont{M.}~\bibnamefont{Mottonen}},
  \bibinfo{author}{\bibfnamefont{V.}~\bibnamefont{Bergholm}},
  \bibinfo{author}{\bibfnamefont{O.-P.} \bibnamefont{Saira}},
  \bibinfo{author}{\bibfnamefont{J.}~\bibnamefont{Zhang}}, \bibnamefont{and}
  \bibinfo{author}{\bibfnamefont{K.~B.} \bibnamefont{Whaley}},
  \bibinfo{journal}{Phys. Rev. A} \textbf{\bibinfo{volume}{77}},
  \bibinfo{pages}{032334} (\bibinfo{year}{2008}).

\bibitem[{\citenamefont{Lee et~al.}(2008)\citenamefont{Lee, Witzel, and
  Sarma}}]{Lee2008}
\bibinfo{author}{\bibfnamefont{B.}~\bibnamefont{Lee}},
  \bibinfo{author}{\bibfnamefont{W.}~\bibnamefont{Witzel}}, \bibnamefont{and}
  \bibinfo{author}{\bibfnamefont{S.~D.} \bibnamefont{Sarma}},
  \bibinfo{journal}{Phys. Rev. Lett.} \textbf{\bibinfo{volume}{100}},
  \bibinfo{pages}{160505} (\bibinfo{year}{2008}).

\bibitem[{\citenamefont{Nakamura et~al.}(2002)\citenamefont{Nakamura, Pashkin,
  Yamamoto, , and Tsai}}]{Nakamura2002}
\bibinfo{author}{\bibfnamefont{Y.}~\bibnamefont{Nakamura}},
  \bibinfo{author}{\bibfnamefont{Y.}~\bibnamefont{Pashkin}},
  \bibinfo{author}{\bibfnamefont{T.}~\bibnamefont{Yamamoto}}, ,
  \bibnamefont{and} \bibinfo{author}{\bibfnamefont{J.}~\bibnamefont{Tsai}},
  \bibinfo{journal}{Phys. Rev. Lett.} \textbf{\bibinfo{volume}{88}},
  \bibinfo{pages}{047901} (\bibinfo{year}{2002}).

\bibitem[{\citenamefont{Faoro and Ioffe}(2006)}]{Faoro2006}
\bibinfo{author}{\bibfnamefont{L.}~\bibnamefont{Faoro}} \bibnamefont{and}
  \bibinfo{author}{\bibfnamefont{L.}~\bibnamefont{Ioffe}},
  \bibinfo{journal}{Phys. Rev. Lett.} \textbf{\bibinfo{volume}{96}},
  \bibinfo{pages}{047001} (\bibinfo{year}{2006}).

\bibitem[{\citenamefont{Koch et~al.}(2007)\citenamefont{Koch, DiVincenzo, and
  Clarke}}]{Koch2007}
\bibinfo{author}{\bibfnamefont{R.}~\bibnamefont{Koch}},
  \bibinfo{author}{\bibfnamefont{D.}~\bibnamefont{DiVincenzo}},
  \bibnamefont{and} \bibinfo{author}{\bibfnamefont{J.}~\bibnamefont{Clarke}},
  \bibinfo{journal}{Phys. Rev. Lett.} \textbf{\bibinfo{volume}{98}},
  \bibinfo{pages}{267003} (\bibinfo{year}{2007}).

\bibitem[{\citenamefont{Leggett et~al.}(1987)\citenamefont{Leggett,
  Chakravarty, Dorsey, Fisher, Garg, and Zwerger}}]{Leggett1987}
\bibinfo{author}{\bibfnamefont{A.~J.} \bibnamefont{Leggett}},
  \bibinfo{author}{\bibfnamefont{S.}~\bibnamefont{Chakravarty}},
  \bibinfo{author}{\bibfnamefont{A.}~\bibnamefont{Dorsey}},
  \bibinfo{author}{\bibfnamefont{M.}~\bibnamefont{Fisher}},
  \bibinfo{author}{\bibfnamefont{A.}~\bibnamefont{Garg}}, \bibnamefont{and}
  \bibinfo{author}{\bibfnamefont{W.}~\bibnamefont{Zwerger}},
  \bibinfo{journal}{Rev. Mod. Phys.} \textbf{\bibinfo{volume}{59}},
  \bibinfo{pages}{1} (\bibinfo{year}{1987}).

\bibitem[{\citenamefont{Uhrig}(2008{\natexlab{b}})}]{Uhrig2008_2}
\bibinfo{author}{\bibfnamefont{G.}~\bibnamefont{Uhrig}},
  \bibinfo{journal}{arxiv.org} p. \bibinfo{pages}{0810.5616}
  (\bibinfo{year}{2008}{\natexlab{b}}).

\bibitem[{\citenamefont{Taylor and Calarco}(2007)}]{Taylor2007}
\bibinfo{author}{\bibfnamefont{J.}~\bibnamefont{Taylor}} \bibnamefont{and}
  \bibinfo{author}{\bibfnamefont{T.}~\bibnamefont{Calarco}},
  \bibinfo{journal}{arxiv.org} p. \bibinfo{pages}{0706.1951}
  (\bibinfo{year}{2007}).

\bibitem[{\citenamefont{Itano et~al.}(1998)\citenamefont{Itano, Bollinger, Tan,
  Jelenkovi\'{c}, Huang, and Wineland}}]{itaw98}
\bibinfo{author}{\bibfnamefont{W.~M.} \bibnamefont{Itano}},
  \bibinfo{author}{\bibfnamefont{J.~J.} \bibnamefont{Bollinger}},
  \bibinfo{author}{\bibfnamefont{J.~N.} \bibnamefont{Tan}},
  \bibinfo{author}{\bibfnamefont{B.}~\bibnamefont{Jelenkovi\'{c}}},
  \bibinfo{author}{\bibfnamefont{X.-P.} \bibnamefont{Huang}}, \bibnamefont{and}
  \bibinfo{author}{\bibfnamefont{D.~J.} \bibnamefont{Wineland}},
  \textbf{\bibinfo{volume}{279}}, \bibinfo{pages}{686} (\bibinfo{year}{1998}).

\bibitem[{\citenamefont{Mitchell et~al.}(1998)\citenamefont{Mitchell,
  Bollinger, Dubin, Huang, Itano, and Baughman}}]{mitchell98}
\bibinfo{author}{\bibfnamefont{T.}~\bibnamefont{Mitchell}},
  \bibinfo{author}{\bibfnamefont{J.}~\bibnamefont{Bollinger}},
  \bibinfo{author}{\bibfnamefont{D.}~\bibnamefont{Dubin}},
  \bibinfo{author}{\bibfnamefont{X.}~\bibnamefont{Huang}},
  \bibinfo{author}{\bibfnamefont{W.}~\bibnamefont{Itano}}, \bibnamefont{and}
  \bibinfo{author}{\bibfnamefont{R.}~\bibnamefont{Baughman}},
  \bibinfo{journal}{Science} \textbf{\bibinfo{volume}{282}},
  \bibinfo{pages}{1290} (\bibinfo{year}{1998}).

\bibitem[{\citenamefont{Huang et~al.}(1998)\citenamefont{Huang, Bollinger,
  Mitchell, and Itano}}]{huap98a}
\bibinfo{author}{\bibfnamefont{X.-P.} \bibnamefont{Huang}},
  \bibinfo{author}{\bibfnamefont{J.~J.} \bibnamefont{Bollinger}},
  \bibinfo{author}{\bibfnamefont{T.~B.} \bibnamefont{Mitchell}},
  \bibnamefont{and} \bibinfo{author}{\bibfnamefont{W.~M.} \bibnamefont{Itano}},
  \bibinfo{journal}{Phys. Rev. Lett.} \textbf{\bibinfo{volume}{80}},
  \bibinfo{pages}{73} (\bibinfo{year}{1998}).

\bibitem[{\citenamefont{Jensen et~al.}(2005)\citenamefont{Jensen, Hasegawa,
  Bollinger, and Dubin}}]{jenm05}
\bibinfo{author}{\bibfnamefont{M.~J.} \bibnamefont{Jensen}},
  \bibinfo{author}{\bibfnamefont{T.}~\bibnamefont{Hasegawa}},
  \bibinfo{author}{\bibfnamefont{J.~J.} \bibnamefont{Bollinger}},
  \bibnamefont{and} \bibinfo{author}{\bibfnamefont{D.~H.~E.}
  \bibnamefont{Dubin}}, \bibinfo{journal}{Phys. Rev. Lett.}
  \textbf{\bibinfo{volume}{94}}, \bibinfo{pages}{025001}
  (\bibinfo{year}{2005}).

\bibitem[{\citenamefont{Brewer et~al.}(1988)\citenamefont{Brewer, Prestage,
  Bollinger, Itano, Larson, and Wineland}}]{brel88}
\bibinfo{author}{\bibfnamefont{L.~R.} \bibnamefont{Brewer}},
  \bibinfo{author}{\bibfnamefont{J.~D.} \bibnamefont{Prestage}},
  \bibinfo{author}{\bibfnamefont{J.~J.} \bibnamefont{Bollinger}},
  \bibinfo{author}{\bibfnamefont{W.~M.} \bibnamefont{Itano}},
  \bibinfo{author}{\bibfnamefont{D.~J.} \bibnamefont{Larson}},
  \bibnamefont{and} \bibinfo{author}{\bibfnamefont{D.~J.}
  \bibnamefont{Wineland}}, \bibinfo{journal}{Phys. Rev. A}
  \textbf{\bibinfo{volume}{38}}, \bibinfo{pages}{859} (\bibinfo{year}{1988}).

\bibitem[{\citenamefont{Itano et~al.}(1993)\citenamefont{Itano, Bergquist,
  Bollinger, Gilligan, Heinzen, Moore, Raizen, and Wineland}}]{itaw93}
\bibinfo{author}{\bibfnamefont{W.~M.} \bibnamefont{Itano}},
  \bibinfo{author}{\bibfnamefont{J.~C.} \bibnamefont{Bergquist}},
  \bibinfo{author}{\bibfnamefont{J.~J.} \bibnamefont{Bollinger}},
  \bibinfo{author}{\bibfnamefont{J.~M.} \bibnamefont{Gilligan}},
  \bibinfo{author}{\bibfnamefont{D.~J.} \bibnamefont{Heinzen}},
  \bibinfo{author}{\bibfnamefont{F.~L.} \bibnamefont{Moore}},
  \bibinfo{author}{\bibfnamefont{M.~G.} \bibnamefont{Raizen}},
  \bibnamefont{and} \bibinfo{author}{\bibfnamefont{D.~J.}
  \bibnamefont{Wineland}}, \bibinfo{journal}{Phys. Rev. A}
  \textbf{\bibinfo{volume}{47}}, \bibinfo{pages}{3554} (\bibinfo{year}{1993}).

\bibitem[{\citenamefont{Pasini et~al.}(2008)\citenamefont{Pasini, Fischer,
  Karbach, and Uhrig}}]{Uhrig2008_3}
\bibinfo{author}{\bibfnamefont{S.}~\bibnamefont{Pasini}},
  \bibinfo{author}{\bibfnamefont{T.}~\bibnamefont{Fischer}},
  \bibinfo{author}{\bibfnamefont{P.}~\bibnamefont{Karbach}}, \bibnamefont{and}
  \bibinfo{author}{\bibfnamefont{G.}~\bibnamefont{Uhrig}},
  \bibinfo{journal}{Phys. Rev. A} \textbf{\bibinfo{volume}{77}},
  \bibinfo{pages}{032315} (\bibinfo{year}{2008}).

\bibitem[{\citenamefont{Weinacht and Bucksbaum}(1999)}]{Weinacht99}
\bibinfo{author}{\bibfnamefont{T.}~\bibnamefont{Weinacht}} \bibnamefont{and}
  \bibinfo{author}{\bibfnamefont{P.}~\bibnamefont{Bucksbaum}},
  \bibinfo{journal}{Nature} \textbf{\bibinfo{volume}{397}},
  \bibinfo{pages}{233} (\bibinfo{year}{1999}).

\end{thebibliography}
\end{document}